\documentclass[conference]{IEEEtran}

\ifCLASSINFOpdf
\else
\fi
\usepackage{tikz}
\usepackage{bbm}
\usepackage{booktabs}

\usepackage{algorithmic}

\usepackage{color}
\usepackage[cmex10]{amsmath}
\usepackage{amssymb}

\begin{document}
%
\title{An Improved Node Ranking for Label Propagation and Modularity based Clustering}

\author{\IEEEauthorblockN{Alessandro Berti}
\IEEEauthorblockA{Department of Mathematics\\
University of Padova\\
35121 Padova\\
Email: berti@math.unipd.it}}


%


\maketitle

\begin{abstract}
In this paper I'll speak about non-spectral clustering techniques and see how a node ordering based
on centrality measures can improve the quality of communities detected. I'll also discuss an
improvement to existing techniques, which further improves modularity.
\end{abstract}


%
\IEEEpeerreviewmaketitle

\section{Introduction}

There is a growing interest, in community detection, for non-spectral clustering techniques.
These don't require to work with matrices (i.e. they don't have to find eigenvalues \cite{spectral}, or to
find a Non-Negative Matrix factorization \cite{nmf}), so they are eventually simpler to implement,
and can be applied to larger graphs. These methods require the nodes to be visited iteratively in a certain order.
While for LP algorithm \cite{labelpropagation} the effect of ordering has been considered (at least for
unweighed graphs \cite{xing2014node}), for Modularity based approaches (like \cite{multilevel}) the interest on initial node ordering
has been low. In this paper I'll show how the choice of an appropriate ordering can improve a lot the quality of the communities
detected.

\section{Background}

I'll consider in this paper undirected weighted graphs $G = (V,E,W)$ where $V$ are the vertices, $E$ the edges and $W$ are the (non-negative) weights associated to the edges.

(Crisp) clustering is the task of assigning each node to a single cluster / community.
A very fast non-spectral method is Label Propagation algorithm (LP, \cite{labelpropagation}), which is an iterative
method visiting nodes in a given order and assigning them the ``label'' which has the greatest appearance in the neighbourhood of the node.
If there are two labels with the same weight, one of them is randomly chosen. The method stops when the label of each node is stabilised
to a value.
Another method is based on Modularity maximization. Modularity (see \cite{multilevel}) is a quantity that takes in account
both the internal cohesion and the separation of the found clusters, and is a widely adopted approach in judging quality of a clustering.
Finding the global maximum of modularity is an hard task, infeasible for large
graphs, but finding a good local maximum can be done in a near linear time, using the algorithm described in \cite{multilevel}, which I'll refer as
Multilevel algorithm. This is an agglomerative method, starting from each node forming a distinct cluster, which visits nodes in a given order, taking
them away from their cluster and putting them in the most appropriate cluster, in order to maximize modularity.
So, both methods considered \cite{labelpropagation,multilevel} take in account the ordering of nodes.

An effective way to order nodes is according to a centrality measure \cite{centrality} on the graph. There are various centrality measures,
which are useful to understand how much a node $v$ is important in the description of the community structure:
\begin{itemize}
\item {\bf Weighted ``degree'' centrality} ({\it deg}): A very simple measure that is the sum of the weights of the edges adjacent to $v$.
\item {\bf PageRank centrality} ({\it pag}): An approach described in \cite{pagerank}.
\item {\bf Closeness centrality} ({\it clo}): Defined as the inverse of the sum of the shortest path between $v$ and other nodes in the graph.
$${\mbox clo}(v) = \frac{1}{\sum_{i \neq v} d(i,v)}$$
$$d(i,v) = \min_{e1,\ldots,e_n \in E, i \in e_1, v \in e_n} \sum_{1 \leq j \leq n} \frac{1}{w_{e_j}}$$
\item {\bf Betweenness centrality} ({\it bet}): Defined as the number of shortest paths between nodes that pass through $v$.
\end{itemize}

Computationally wise, degree centrality is the easiest to compute, followed by PageRank, while closeness centrality is a bit more difficult
as it require to compute the length of the shortest paths. Betweenness centrality also require to compute shortest paths and observe the number of these
passing over a node.

\begin{table*}[t]
\centering
\caption{\label{tbl:lp} Average (on 1000 graphs) Label Propagation clustering modularity with different nodes order. I get the best results with ``my'' centrality measure.}
\begin{tabular}{l|cccccc}
\toprule
{\bf Parameters} & {\bf random ord.} & {\bf deg ord.} & {\bf pag ord.} & {\bf clo ord.} & {\bf bet ord.} & {\bf my ord.} \\
\midrule
n=50, ~nei=4 & 0.0895 & 0.5675 & 0.5677 & 0.5679 & 0.5710 & {\bf 0.5714} \\
n=50, ~nei=5 & 0.0047 & 0.5184 & 0.5201 & 0.5126 & 0.5192 & {\bf 0.5202} \\
n=50, ~nei=6 & -0.0001 & 0.4669 & 0.4681 & 0.4574 & 0.4654 & {\bf 0.4685} \\
n=100, nei=7 & -0.0001 & 0.5964 & 0.5967 & 0.6012 & 0.6010 & {\bf 0.6018} \\
n=100, nei=8 & -0.0001 & 0.5716 & 0.5719 & 0.5754 & 0.5729 & {\bf 0.5763} \\
n=100, nei=9 & -0.0001 & 0.5460 & 0.5479 & 0.5449 & 0.5474 & {\bf 0.5510} \\
\bottomrule
\end{tabular}
\end{table*}

\begin{table*}[t]
\centering
\caption{\label{tbl:ml} Average (on 1000 graphs) Multilevel clustering modularity with different nodes order. I get the best results with ``my'' centrality measure. Natural ordering means the order in which nodes were added to the graph.}
\begin{tabular}{l|cccccc}
\toprule
{\bf Parameters} & {\bf natural ord.} & {\bf deg ord.} & {\bf pag ord.} & {\bf clo ord.} & {\bf bet ord.} & {\bf my ord.} \\
\midrule
n=50, ~nei=4 & 0.5841 & 0.5880 & {\bf 0.5886} & 0.5881 & 0.5884 & 0.5885 \\
n=50, ~nei=5 & 0.5312 & 0.5376 & 0.5373 & 0.5371 & 0.5375 & {\bf 0.5377} \\
n=50, ~nei=6 & 0.4869 & 0.4937 & 0.4933 & 0.4936 & 0.4933 & {\bf 0.4938} \\
n=100, nei=7 & 0.5910 & 0.6142 & 0.6144 & {\bf 0.6147} & 0.6137 & {\bf 0.6147} \\
n=100, nei=8 & 0.5599 & 0.5869 & 0.5869 & 0.5870 & 0.5869 & {\bf 0.5873} \\
n=100, nei=9 & 0.5320 & 0.5623 & 0.5623 & 0.5615 & 0.5620 & {\bf 0.5624} \\
\bottomrule
\end{tabular}
\end{table*}

I'll consider the ordering that takes as first the lower centrality nodes, as they are less important in the definition of communities
and can be suddenly merged, by the considered iterative methods \cite{labelpropagation,multilevel}, to high centrality nodes.
If two or more nodes have the same centrality, then the order between them is chosen randomly.
This type of ordering shows, in applications, big gains for each of the considered centrality measures, in comparison to a random ordering.

\section{My method}

In applications, we'll see that closeness and betweenness have the best behaviour. I will try now to merge them, in order to obtain a better ordering.
Firstly, we can note that betweenness is an integer quantity (the number of shortest paths passing through the node) and that is not uncommon that two
nodes share the same betweenness. Now, I introduce a quantity that can be called ``relative'' closeness of a node:
$$C(i) = \frac{c(i)}{2 \cdot \max_{v \in G} c(v) + 1}$$
The above quantity is surely contained in the interval $[0,1/2)$. The idea is to define the quantity
$$\mbox{my}(v) = \mbox{bet}(v) + C(v)$$
and doing the ordering according to it. We can see that if $\mbox{bet}(v) < \mbox{bet}(w)$ then $\mbox{my}(v) < \mbox{my}(w)$, so the only effect
of this measure is to do a correct ordering in the cases where $\mbox{bet}(v) = \mbox{bet}(w)$, which are indeed critical.

\section{Assessment}

The assessment has been done on Watts-Strogatz graphs, all generated with parameters p $= 0$ (rewiring probability) and dim $= 1$
(dimension of the starting lattice).
For each of the considered methods (Label Propagation \cite{labelpropagation} and Multilevel \cite{multilevel}), and each choice of the parameters
(n is the number of nodes, while nei is the neighborhood within which the vertices of the lattice will be connected) I have calculated
the average modularity (on $1000$ graphs), by ordering the nodes in the indicated ways, of the clustering, and reported the results
in Table \ref{tbl:lp} and Table \ref{tbl:ml}.

The biggest gains are produced by ordering based on betweenness and closeness measures. Initially avoiding nodes with high betweenness is
indeed related to treating later possible overlapping nodes, which are between communities and risk to ruin clustering. Closeness is important
because nodes with low closeness are farther from the others and so have less influence, and taking in account the distance, instead of the ``direct'' arc, is
more meaningful.

\section{Conclusions}

In this paper I have examined how, ordering the nodes according to a centrality measure, I can get better quality of the clustering produced, according
to modularity. I have also introduced a measure that, while can't be called ``novel'' approach, has shown good behaviour in applications. Although not obtaining
the best results, calculation of degree centrality is very easy and I suggest it as a plausible pre-processing step, without ruining the performance at all,
to easily improve the modularity.

\section{Acknowledgements}

This work has been supported by FSE fellowship 2105/201/17/1148/2013.

\bibliographystyle{IEEEtran}
\bibliography{ranking}
\end{document}